\documentstyle[preprint,aps,psfig,bezier]{revtex}
\begin{document}
\draft
\tighten
\preprint{}

\title{Towards exact results of QED from supersymmetry} 

\author{Masato Arai and Noriaki Kitazawa}
\address{Department of Physics, Tokyo Metropolitan University \\
Hachioji, Tokyo 192-0397, Japan}

\date{\today}

\preprint{
\parbox{5cm}{
TMUP-HEL-9903\\
}}

\maketitle
\begin{abstract}
To obtain some exact results of U(1) gauge theory (QED), we construct the
 low energy effective action of {\it N}=2 supersymmetric QED with a
 massless matter and Fayet-Iliopoulos term, assuming no
 confinement. The harmonic superspace formalism for {\it N}=2 extended
 supersymmetry makes the construction easy. 
 We analyze
 the vacuum structure and find no vacuum.
 It suggests the confinement in non-supersymmetric QED at low energies.
\end{abstract}
\newpage
\section{Introduction}
QED is the most successful quantum field theory in the phenomenological
point of view. In fact, QED perfectly describes the
electromagnetic interaction at low energies. However, there is a question
whether QED is a fully
consistent theory beyond the perturbation theory.
QED has the Landau ghost problem\cite{landau} and the renormalized
coupling constant vanishes. 
This means QED may be trivial as a quantum field theory, and
it can only be regarded as a low energy effective theory.

On the other hand, Miransky suggested that QED is 
non-trivial\cite{Miransky}. He investigated a truncated
Schwinger-Dyson equation for the fermion propagator and found a
continuous chiral
phase transition. He claimed that the chiral symmetry is spontaneously
broken in the strong coupling phase. After his work, non-perturbative studies of QED have
been done extensively\cite{leung,Kondo,yamawaki}.
Some of numerical simulations were carried out to understand whether QED is
trivial or not. Kogut et al. claimed that the
existence of a chiral phase transition was confirmed by numerical
studies\cite{kogut}. On the other hand, DESY-J{\"u}lich group claimed that QED is a
trivial theory which is described by a Gaussian fixed point, and the critical
behavior around it is similar to the one of the $\lambda\phi^4$
model\cite{DESY}. This controversy is not resolved yet.

Recently, there has been much progress in the understanding of the non-perturbative
dynamics of {\it N}=1 and {\it N}=2 supersymmetric four-dimension field
theories. The exact superpotential can be derived 
in {\it N}=1 supersymmetric QCD (SQCD: supersymmetric SU$(N_c)$ gauge
theory with vector-like matters) \cite{Seiberg}, and the models with
various gauge symmetries and matter contents have been investigated. Seiberg and Witten derived
the exact low energy effective action for {\it N}=2 supersymmetric SU(2) Yang-Mills theory in Coulomb phase up to two
derivatives\cite{s-w1}, 
and generalized it to the case of {\it N}=2 SQCD \cite{s-w2}. 
Their method was applied to the different gauge groups and the solution was obtained. 

Since we can derive the exact low energy effective action (LEEA) of
{\it N}=2 supersymmetric  gauge theories,
we can expect to extract the exact information of non-supersymmetric
gauge theories, QED and QCD, for example. 
A simple way to break supersymmetry is to add soft supersymmetry breaking terms. 
In Refs. \cite{masiero,peskin,sakai,hsu,martin,marino} soft breaking terms are used to explore {\it N}=1 supersymmetric
QCD and the phase structure of these theories in the absence of
supersymmetry. 
We will focus on {\it N}=2 supersymmetric QED (SQED) with a massless matter to
explore {\it N}=0 QED. It is well known that 
Fayet-Iliopoulos (FI)
term spontaneously breaks supersymmetry in {\it N}=2 SQED \cite{fayet,tree,Ivanov}. We
construct the exact LEEA of SQED with FI
term not introducing the soft breaking terms by hand. 

In 
Refs. \cite{s-w1,s-w2}\cite{masiero,peskin,sakai,hsu,martin,marino} {\it N}=1 superfields were
used to describe {\it N}=2 supersymmetric theories. Therefore, {\it N}=2 
supersymmetry was not manifest in those works. We can use constrained superfields on the 
standard {\it N}=2 superspace \cite{grimm}, but these are not
appropriate to the construction and the analysis of the LEEA, because
the description becomes extremely complicated when the
interaction is included on.
An elegant off-shell formulation of {\it N}=2 supersymmetry is the harmonic superspace
formalism, developed by Galperin et al. \cite{Galperin}. In this formalism
superfileds are unconstrained and we do not need to solve complicated
constraints. {\it N}=2 supersymmetry is manifest at each step of the
calculation. We will see that this formalism is very powerful for
constructing the LEEA in this paper.

The plan of this paper is as follows. In Sec.II we briefly review the
harmonic superspace formalism stressing some important points for our main. In Sec.III we construct the LEEA
of SQED without the FI term as the first step. In Sec.IV we extend the
discussion in the 
previous section to the case with FI term. In Sec.V we analyze
the effective potential of the LEEA which is obtained in the previous
section, and discuss the vacuum structure of N=0 QED. Sec.VI is
devoted to the conclusion. Our notations and conventions are
summarized in Appendix.

\newcommand{\ptheta}{\theta^+}
\newcommand{\bptheta}{{\bar{\theta}}^+}
\section{Harmonic Superspace Formalism}
We briefly review some of the basics of the harmonic superspace
formalism (HSS).
HSS is the formalism for {\it N}=2 extended
supersymmetry developed by Galperin et al. \cite{Galperin}. The
standard {\it N}=2 superspace is parameterized by the coordinates
\begin{eqnarray}
 \{x^\mu, \theta_{\alpha i},{\bar{\theta}}_{\dot{\alpha}}^i \},
\end{eqnarray}
where $\alpha$ is the spinor index and  $i$ is SU(2$)_{{\rm R}}$ index.
The key ingredient in HSS is the harmonic variables
$u_i^\pm$ which parameterize the coset space SU(2$)_{{\rm
R}}$/U(1). The variables satisfy the relation
\begin{equation}
u^{+i}u_i^-=1,
\end{equation}
where $\pm$ denote U(1) charge $\pm 1$. The variables of the harmonic
superspace in the central basis (CB) are 
\begin{equation}
{\bf CB}:\{{x^\mu,\theta_{\alpha i},{\bar{\theta}}_{\dot{\alpha}}^i,u_i^\pm}\}.
\end{equation}
Harmonic superfields are the functions of these variables. In CB the
differentiation by the harmonic variables are defined as
\begin{eqnarray}
&D^{++}=u^{+i}\displaystyle\frac{\partial}{\partial u^{-i}},\hspace{2mm}
D^{--}=u^{-i}\displaystyle\frac{\partial}{\partial u^{+i}},&
\end{eqnarray}
and the integration over $u$ is defined by the following rules:
\begin{eqnarray}
&\displaystyle\int du\hspace{1mm}1 = 1,&\\
&\displaystyle\int du\hspace{1mm}u_{(i_1}^+\cdots u_{i_n}^+u_{j_1}^-\cdots u_{j_m)}^-=0,\hspace{3mm}n+m>0,
\end{eqnarray}
where the parenthesis mean symmetrization of SU(2$)_{{\rm R}}$
indices. Namely, the $u$
integration is defined to pick up the SU(2$)_{{\rm R}}$ singlet
part. The Lagrangian which is described by the harmonic superfields is
not manifestly real under the usual complex conjugation. However, it is
real under the conjugation which is the combination of the usual complex
conjugation and the star conjugation. The star conjugation for the harmonic
variables are defined by
\begin{eqnarray}
 (u_i^+)^*=u_i^-,\hspace{2mm}(u_i^-)^*=-u_i^+,
\end{eqnarray}
and other quantities are singlet under the conjugation. The harmonic
variables are transformed under the combined conjugation as
\begin{eqnarray}
\displaystyle {\overline{u^{\pm i}}}^*=-u_i^\pm,\hspace{2mm}{\overline{u_i^\pm}}^*=u^{\pm i}.
\end{eqnarray}

There is another important basis called the analytic basis (AB):
\begin{eqnarray}
{\bf AB}&:&\{x^\mu_A,\theta_\alpha^\pm,{\bar{\theta}}_{\dot{\alpha}}^\pm,u_i^\pm\},\\
&&x^\mu_A=x^\mu-2i\theta^{(i}\sigma^\mu{\bar{\theta}^{j)}}u_i^+u_j^-, \nonumber\\
&&\theta^\pm=\theta^i u_i^\pm,\hspace{3mm}{\bar{\theta}}^\pm={\bar{\theta}}^i u_i^\pm. \nonumber
\end{eqnarray}
Irreducible harmonic 
superfields are not the function of the entire variables of AB or CB but
the function on their subspaces, the analytic subspace (ASS) or the
chiral subspace (CSS). ASS is defined by
\begin{eqnarray}
{\bf ASS}:\zeta_A&=&\{x^\mu_A,\theta^+_\alpha,{\bar{\theta}}^+_{\dot{\alpha}},u^\pm_i\},
\end{eqnarray}
and it is an invariant subspace under {\it N}=2 supersymmetry
transformation. This fact allows one to define the
analytic superfields which satisfy the analyticity conditions
\begin{eqnarray}
D^+\phi^{(q)}(\zeta_A)={\bar{D}}^+\phi^{(q)}(\zeta_A)=0,
\end{eqnarray}
where 
\begin{eqnarray}
 D^+=D^i u_i^+=\displaystyle\frac{\partial}{\partial \theta^-},\hspace{2mm}
{\bar{D}}^+={\bar{D}}^i u_i^+=\frac{\partial}{\partial {\bar{\theta}}^-},
\end{eqnarray}
and 
$q$ denotes U(1) charge of the field. 

There are two basic supermultiplets in the {\it N}=2 supersymmetry: the 
hypermultiplet and the vectormultiplet. Fayet-Sohnius(FS) superfield
\cite{fs} describes the 
complex hypermultiplet whose on-shell physical components are
$(f^i,\psi,\varphi)$, where $f^i$
is a complex scalar in SU(2$)_{{\rm R}}$ doublet, and $(\psi$, $\varphi)$ is a
Dirac spinor\footnote{There is another harmonic superfield,
Howe-Stelle-Townsend superfield, which describes real
hypermultiplet\cite{hst}.}. The superfield with U(1) charge $+1$ is written
down as
\begin{eqnarray}
\phi^{+}(\zeta_A)&=&F^{+}(x_A,u^\pm)+\sqrt{2}\theta^+\psi(x_A,u^\pm)
+\sqrt{2}{\bptheta}{\bar{\varphi}}(x_A,u^\pm)\nonumber\\
&+&\ptheta\ptheta M^{-}(x_A,u^\pm)+\bptheta\bptheta N^{-}(x_A,u^\pm)\nonumber \\
&+&\ptheta\sigma^\mu\bptheta V_\mu(x_A,u^\pm)
+\sqrt{2}\bptheta\bptheta\ptheta\xi_\alpha^{--}(x_A,u^\pm)\nonumber \\
&+&\sqrt{2}\bptheta{\bar{\chi}}^{--}(x_A,u^\pm)
+\ptheta\ptheta\bptheta\bptheta D^{---}(x_A,u^\pm),
\end{eqnarray}
where $F^+,M^-,N^-$ and $D^{---}$ are complex scalar fields,
$\psi,{\bar{\varphi}},\xi^{--}$ and ${\bar{\chi}}^{--}$ are Weyl fermion fields and $V_\mu$ is a complex 
vector field.
Each component field can be expanded in $u^\pm_i$. For example, 
\begin{eqnarray}
F^{+}(x_A,u^\pm)=\displaystyle\sum_{n=0}^\infty f^{(i_1\cdots i_{n+1}j_1\cdots j_n)}(x_A)u^+_{(i_1}\cdots u^+_{i_{n+1}}u^-_{j_1}\cdots u^-_{j_n)}.
\end{eqnarray}
Therefore, FS superfield includes infinite number of auxiliary fields.
The action for a free complex FS hypermultiplet
is given by
\begin{eqnarray}
S_{{\rm FS}}=\displaystyle\int d\zeta_A^{(-4)}du \hspace{1mm}{\overline{\phi^+}}^*D^{++}\phi^+,\label{eq:hm}
\end{eqnarray}
where $D^{++}$ is a covariant derivative in AB given by Eq. (\ref{eq:cov})
and $d\zeta_A^{(-4)}$ is the analytic measure defined by
\begin{eqnarray}
d\zeta_A^{(-4)}=d^4 x_A d^2\ptheta d^2\bptheta .
\end{eqnarray}
Solving the equation of motion $D^{++}\phi^+=0$, we can easily check
that only the physical components $(f^i,\psi,\varphi)$ remain and follow
free equation of motions.

The on-shell physical components of a vectormultiplet are
$(A,A_\mu,\lambda^i)$, where $A$ is a complex scalar, $A_\mu$
is a vector field and  $\lambda^i$ is a
Majorana spinor in SU(2$)_{{\rm R}}$ doublet. 
A vectormultiplet is described by the dimensionless analytic superfield $V^{++}$
of U(1) charge +2. It
transforms under gauge transformation as
\begin{eqnarray}
\delta V^{++}=-D^{++}\lambda(x_A,u^\pm)\label{eq:gtrans}
\end{eqnarray}
in abelian case, where $\lambda$ is an analytic superfield with U(1)
charge $0$.
Here $V^{++}$ is chosen to be real,
namely,
\begin{eqnarray}
V^{++}={\overline{V^{++}}}^*.
\end{eqnarray}
If we take the Wess-Zumino-like gauge,
\begin{eqnarray}
V^{++}(\zeta_A)&=&\ptheta\ptheta\frac{1}{\sqrt{2}}A(x_A,u^\pm)+\bptheta\bptheta\frac{1}{\sqrt{2}}A^*(x_A,u^\pm)+i\ptheta\sigma^\mu\bptheta A_\mu(x_A,u^\pm) \nonumber\\
&&+\bptheta\bptheta\ptheta 2\lambda^-(x_A,u^\pm)
+\ptheta\ptheta\bptheta 2{\bar{\lambda}}^-(x_A,u^\pm)
+\ptheta\ptheta\bptheta\bptheta D^{--}(x_A,u^\pm),
\end{eqnarray}
where $D^{--}=D^{(ij)}u_i^- u_j^-$ is a real auxiliary field in
SU(2$)_{{\rm R}}$ triplet.

CSS is defined by
\begin{eqnarray}
{\bf CSS}:\zeta_R&=&\{x^\mu_R,\theta_{\alpha i},u_i^\pm\},\\
&&x^\mu_R=x^\mu-i\theta^i\sigma^\mu{\bar{\theta}}_i, \nonumber
\end{eqnarray}
and the gauge field strength superfield $W$
is described as a function on it.
\begin{eqnarray}
W(x_{R},\theta)&=&-\displaystyle\frac{1}{4}\int du ({\bar{D}}^+)^2 V^{++}\\
&=&\displaystyle\frac{1}{\sqrt{2}}A^*(x_R)
-\frac{1}{3\sqrt{2}}\epsilon_{ik}\epsilon_{jl}(\theta^i\theta^j)(\theta^k\theta^l)\Box A(x_R)
-\frac{1}{4}\theta^i\sigma^\mu{\bar{\sigma}}^\nu\theta_i F_{\mu\nu}(x_R) \nonumber \\
&&+\theta^i\lambda_i(x_R)+\frac{2}{3}i(\theta^i\theta^j)(\theta^k\sigma^\mu\partial_\mu{\bar{\lambda}}^l(x_R))\epsilon_{ik}\epsilon_{jk}+\frac{1}{3}\epsilon_{ik}\epsilon_{jk}(\theta^i\theta^j)D^{(kl)}(x_R).\label{eq:fieldst}
\end{eqnarray}
The superfield which is a function on CSS is called the chiral superfield.
Note that the chiral superfield $W$ does not explicitly depend on ${\bar{\theta}}_i$ and
$u_i^\pm$ and can not the function on ASS. Also note that the analytic
superfield can not be described as a function on CSS. The action of a vectormultiplet is
given by
\begin{eqnarray}
S_{{\rm gauge}}=\frac{1}{4\pi}{\rm Im}\tau\int d\zeta_R W^2,
\end{eqnarray}
where $\tau=i\frac{4\pi}{e^2}+\frac{\Theta}{2\pi}$ with that $e$ is the gauge
coupling and $\Theta$ is the vacuum angle. $d\zeta_R=d^4x_R d^4\theta du$ is the 
chiral subspace measure.

We write down the tree-level action of SQED with single matter as
\begin{eqnarray}
S=\displaystyle\int d\zeta_A^{(-4)}du\hspace{1mm}{{\overline{\phi^+}}}^*(D^{++}+2iV^{++})
\phi^+ + \frac{1}{4\pi}{\rm Im}\tau\int d\zeta_R W^2.
\end{eqnarray}
The integrand of the first term (the analytic part) must have U(1) charge
$+4$ and does not explicitly depend on
$\theta^-$ and ${\bar{\theta}}^-$, i.e., it must be
analytic. The chiral superfield does not appear in the analytic
part, because the chiral superfield does not satisfy the analyticity.
Similarly, we find that the analytic superfield does not appear in
the integrand of the second term (the chiral part). The chiral part does
not explicitly depend on ${\bar{\theta}_i}$ and $u_i^\pm$. These facts are important
for constructing the LEEA.

\section{Construction of the LEEA without Fayet-Iliopoulos term}\label{leea1}
In this section we construct the LEEA of SQED with single
massless matter using the harmonic superspace formalism. In the next section
we apply the method used in this section to the case of including FI term. 
The tree-level action leads the scalar potential
\begin{eqnarray}
V=|\sqrt{2}A|^2{\bar{f}}^i f_i+\displaystyle\frac{e_0^2}{2}({\bar{f}}^i f_i)^2,
\end{eqnarray}
where $e_0$ is the bare coupling constant, $A$ and $f^i$ are the complex
scalar fields in the vectormultiplet and FS
 hypermultiplet, respectively. The classical moduli space is
 parameterized by the vacuum expectation value of the complex scalar field $A$. 
 In case of single matter, 
 $f^i$ has no
 vacuum expectation value and the gauge symmetry
 is not broken. Namely, the theory is always in Coulomb phase. If we
 consider multiple matter, the moduli space has Higgs branch in which
 the gauge symmetry is broken.

Our strategy of getting the LEEA is the same which was developed in
Ref. \cite{Seiberg}.  The LEEA must be invariant under the enlarged
symmetry transformation in which the parameters of the theory
transform. These parameters can be considered as the vacuum expectation
values of some external superfields. The holomorphy (or analyticity) also
constrains the LEEA. By using the information obtained in the weak
coupling limit, we can determine the LEEA. 

The transformation laws of the fields and parameters in the fundamental theory is summarized in table \ref{sym1}.

We assume that there is no confiment at low
energies. If the resultant LEEA has no inconsistency, we can conclude
that this assumption is justified.

The general
form of the LEEA of the chiral part (lowest order in the derivative
expansion) is given by
\begin{eqnarray}
{\cal L}_{{\rm C}}=\displaystyle\frac{1}{4\pi}{\rm Im}\int d^4\theta g(W,\Lambda),\label{eq:gauge1}
\end{eqnarray}
where $g(W,\Lambda)$ is a holomorphic function which satisfies
the following conditions.
\begin{enumerate}
 \item U(1) charge $0$.
 \item mass dimension $2$.
 \item U(1$)_{{\rm R}}$ charge 4.
 \item gauge singlet.
\end{enumerate}
We stress again that
the FS superfield
can not appear in the chiral part. The parameter $\Lambda$ can be
understood as the vacuum expectation value of the lowest component of a
chiral superfield. The above conditions
restrict Eq. (\ref{eq:gauge1}) to be the form
\begin{eqnarray}
{\cal L}_{{\rm C}}=\displaystyle\frac{1}{4\pi}{\rm Im}\int d^4\theta G\left(\frac{\Lambda}{W}\right)W^2.
\end{eqnarray}
We can estimate $G$ at one-loop level in the weak coupling limit $\Lambda\rightarrow \infty$. Namely, we
can get
\begin{eqnarray}
\displaystyle\lim_{\Lambda\rightarrow \infty}G\left(\frac{\Lambda}{W}\right)=\frac{i}{\pi}\ln\frac{\Lambda}{W}.
\end{eqnarray}
Thus we obtain
\begin{eqnarray}
G\left(\frac{\Lambda}{W}\right)=\frac{i}{\pi}\ln\frac{\Lambda}{W}+\tilde{G}\left(\frac{\Lambda}{W}\right),
\end{eqnarray}
where $\tilde{G}$ includes the non-perturbative effect. 
We assume that $\tilde{G}$ does not have
singularities, namely, all massless particles have been already included. Then, the Liouville theorem leads
\begin{eqnarray}
\tilde{G}\left(\frac{\Lambda}{W}\right)={\rm constant}.
\end{eqnarray}
Therefore, the chiral part is determined as
\begin{eqnarray}
{\cal L}_{{\rm C}}=\displaystyle\frac{1}{4\pi}{\rm Im}\int d^4\theta \frac{i}{\pi}W^2\ln\frac{\Lambda}{W}.\label{eq:gauge} 
\end{eqnarray}
This is exactly the same result given by Seiberg and Witten
\cite{s-w2}. Note that the singularity at $\langle W \rangle=0$ is not removed in
spite of considering the elementary matter field. The theory is not
defined at $\langle W \rangle=\langle A \rangle=0$ within our assumptions.

Next, we determine the LEEA of the analytic part. The general
form is given by
\begin{eqnarray}
{\cal L}_{{\rm A}}=\displaystyle\int d^2\ptheta d^2\bptheta du \hspace{1mm}f^{(+4)}(\phi^+,{\overline{\phi^+}}^*,V^{++},{\cal D^{++}}),
\end{eqnarray}
where ${\cal D^{++}}$ represents the covariant derivative ${\cal
D^{++}}=D^{++}+2iV^{++}$. 
Analytic function $f^{(+4)}$ must satisfy the
following conditions.
\begin{enumerate}
 \item U(1) charge 4.
 \item mass dimension 2.
 \item U(1$)_{\rm R}$ charge 0.
 \item gauge singlet.
\end{enumerate} 
We stress again that the chiral superfield $W$ can not appear in
the analytic part. Considering the above conditions, we
obtain
\begin{eqnarray}
{\cal L}_{{\rm FS}}=\displaystyle\int d^2\ptheta d^2\bptheta du \hspace{1mm}{\overline{\phi^+}}^*{\cal D^{++}}\phi^+.
\end{eqnarray}
Surprisingly, this is the same form of the tree-level one.
The first derivation of the LEEA of the hypermultiplet of
SQED and SQCD was done in Ref. \cite{ketov} using the harmonic
superspace formalism. In Ref. \cite{ketov} the self-interaction of the
massive FS hypermultiplet
is derived by the perturbative calculation:
\begin{eqnarray}
 \Delta{\cal L}=\lambda\displaystyle\int d^2\ptheta d^2\bptheta du \hspace{1mm}
({\overline{\phi^+}}^*\phi^+)^2,\end{eqnarray}
where $\lambda$ includes an infrared cutoff. The self-interaction term
does not appear in our method based on the symmetry and holomorphy even
in the massive case. It is expected that the infrared divergence
disappears by summing up all the one-loop diagrams with external FS
superfields, and only the higher order terms in the derivative expansion
are obtained.

The total LEEA of SQED is 
\begin{eqnarray}
S_{{\rm eff}}=\displaystyle\int d\zeta_{A}^{(-4)} du \hspace{1mm}{\overline{\phi^+}}^*{\cal D^{++}}\phi^++\displaystyle\frac{1}{4\pi}{\rm Im}\int d\zeta_R \frac{i}{\pi}W^2\ln\frac{\Lambda}{W}.
\end{eqnarray}
We remark the modification of the moduli space by the quantum effect. 
The quantum effect forbids a part of the moduli space $\langle W
\rangle=\langle A \rangle >\Lambda$
where the effective coupling $\alpha_{{\rm eff}}$ is negative.

\section{Construction of the LEEA with Fayet-Iliopoulos term}
We construct
the LEEA of SQED with spontaneous supersymmetry breaking to get some exact results of
N=0 QED. In case of SQED, we can introduce 
FI term 
\begin{eqnarray}
{\cal L_{{\rm FI}}}=\displaystyle\int d^2\ptheta d^2\bptheta du \hspace{1mm}\xi^{++}V^{++}=\frac{1}{3}\xi^{ij}D_{(ij)}, \hspace{2mm}\xi^{++}\equiv\xi^{ij}u_i^+u_j^+ \label{eq:F.I}
\end{eqnarray}
to break supersymmetry spontaneously, where $\xi^{ij}$ includes three
real parameters $\xi^{(a)}$ of mass dimension
2:
\begin{eqnarray}
\xi&=&i\xi^{(a)}(\sigma^{a}\epsilon)=
 \left(
    \begin{array}{cc}
	i\xi^{(1)}+\xi^{(2)} & -i\xi^{(3)} \\
        -i\xi^{(3)}           & -i\xi^{(1)}+\xi^{(2)}
    \end{array}
    \right). \label{eq:xi}
\end{eqnarray}

The procedure of constructing the LEEA is the same as that in the
previous section. The transformation laws for the fields and parameters are summarized in table
\ref{sym2}. The parameters $\xi^{ij}$ can be understood as the vacuum
expectation value of the analytic superfield $\xi^{++}$.

First we consider the LEEA of the chiral part. Repeating the same arguments in the
previous section, we obtain the general form 
\begin{eqnarray}
{\cal L}_{{\rm C}}=\displaystyle\frac{1}{4\pi}{\rm Im}\int d^4\theta g(W,\Lambda)=\frac{1}{4\pi}{\rm Im}\int d^4\theta G\left(\frac{\Lambda}{W}\right)W^2.
\end{eqnarray}
This is exactly the same form that is obtained in the case without FI
term. The
coefficient $\xi^{++}$ can not be included in $G$. After all, using the
one-loop result for $G$, the LEEA of the chiral part is given by Eq. (\ref{eq:gauge}). 

Next we consider the LEEA of the analytic part. The general form is 
\begin{eqnarray}
{\cal L}_{{\rm A}}=\displaystyle\int d^2\ptheta d^2\bptheta du \hspace{1mm}f\left(\frac{\xi^{++}}{\phi^+{\overline{\phi^+}}^*}\right){\overline{\phi^+}}^*{\cal D^{++}}\phi^+.
\end{eqnarray}
We can estimate the function
$f\left(\frac{\xi^{++}}{\phi^+{\overline{\phi^+}}^*}\right)$ in the weak
coupling limit $\xi^{++}\rightarrow 0$ using the perturbation theory. We 
find that there is no one particle irreducible diagram which includes
$\xi^{++}$ and conclude
\begin{eqnarray}
\displaystyle\lim_{\xi\rightarrow 0} f\left(\frac{\xi^{++}}{\phi^+{\overline{\phi^+}}^*}\right)={\rm constant}.
\end{eqnarray}
We can make the constant unity by rescaling the field
$\phi^+$. Including the non-perturbative effect, $f$ is given by
\begin{eqnarray}
 f\left(\frac{\xi^{++}}{\phi^+{\overline{\phi^+}}^*}\right)=1+\tilde{f}\left(\frac{\xi^{++}}{\phi^+{\overline{\phi^+}}^*}\right),
\end{eqnarray}
where $\tilde{f}$ describes non-perturbative effect. 
Here, we assume again that all massless fields have been already included and 
the analytic function has no singularity.
The Liouville theorem leads
\begin{eqnarray}
 f\left(\frac{\xi^{++}}{\phi^+{\overline{\phi^+}}^*}\right)=1.
\end{eqnarray}
Therefore, after all, the LEEA of the analytic part is exactly the same
with that is obtained in the case without FI term.

The FI term in Eq. (\ref{eq:F.I}) is the exact form.
An analytic function $h\left(\frac{\xi^{++}}{\phi^+{\overline{\phi^+}}^*}\right)$ seems to be allowed as the
coefficient function of FI term. However the function must be a constant 
due to the gauge invariance. Note that $V^{++}$ is gauge invariant up to 
the total derivative.

We conclude that the LEEA of SQED with FI term is given by 
\begin{eqnarray}
S_{{\rm eff}}=\displaystyle\int d \zeta_A^{(-4)} du \hspace{1mm}{\overline{\phi^+}}^*{\cal D^{++}}\phi^++\displaystyle\frac{1}{4\pi}{\rm Im}\int d\zeta_R \frac{i}{\pi}W^2\ln\frac{\Lambda}{W}+\displaystyle\int d\zeta_A^{(-4)} du \hspace{1mm}\xi^{++}V^{++}.
\end{eqnarray}
\section{Potential analysis of N=0 QED }\label{potesec}
In this section, we write down and analyze the effective potential of the LEEA
which is obtained in the previous section. 
We take the polar
decomposition
\begin{eqnarray}
A=a e^{i\sigma},
\end{eqnarray}
where $a$ and $\sigma$ are real scalar fields. 
The contribution to the potential from the analytic part including FI
term is
\begin{eqnarray}
V_{{\rm A}}=(\sqrt{2}a)^2{\bar{f}}^i f_i+\displaystyle\frac{2}{3}i{\bar{f}}^i f^j D_{(ij)}-\frac{1}{3}\xi^{ij}D_{(ij)}.
\end{eqnarray}
The contribution from the chiral part is
\begin{eqnarray}
V_{{\rm C}}=\displaystyle\frac{1}{8\pi^2}\left(\frac{1}{9}D^{(ij)}D_{(ij)}\ln\frac{\Lambda}{a}-\frac{1}{6}D^{(ij)}D_{(ij)}+{\rm h.c}\right).
\end{eqnarray}
Using the equation of motion of the auxiliary field $D^{(ij)}$, we obtain
the total scalar potential as
\begin{eqnarray}
V_{{\rm eff}}&=&V_{{\rm A}}+V_{{\rm C}}\nonumber\\
&=&(\sqrt{2}a)^2{\bar{f}}^i f_i-\displaystyle\frac{4\pi^2}{\ln\frac{\Lambda^\prime}{a}}\left({\bar{f}}^if^j+\frac{i}{2}\xi^{ij}\right)\left({\bar{f}}_if_j+\frac{i}{2}\xi_{ij}\right),
\end{eqnarray}
where $\Lambda^\prime=\Lambda e^{-3/2}$.
Note that the potential is independent of the scalar field $\sigma$. The 
vacuum expectation value of $\sigma$ is unphysical, since $\Theta$ term
in U(1) gauge theory has no meaning. 
The extremal conditions for $a$ and $f$ are
\begin{eqnarray}
\displaystyle\frac{\partial V_{{\rm eff}}}{\partial {\bar{f}}^i}&=&\left\{(\sqrt{2}a)^2\epsilon_{ij}-\displaystyle\frac{8\pi^2}{\ln\frac{\Lambda^\prime}{a}}\left({\bar{f}}_if_j+\frac{i}{2}\xi_{ij}\right)\right\}f^j=0,\\
\displaystyle\frac{\partial V_{{\rm eff}}}{\partial a}&=&2a{\bar{f}}^if_i+\frac{1}{a}\displaystyle\frac{4\pi^2}{(\ln\frac{\Lambda^\prime}{a})^2}\left({\bar{f}}^if^j+\frac{i}{2}\xi^{ij}\right)\left({\bar{f}}_if_j+\frac{i}{2}\xi_{ij}\right)=0.
\end{eqnarray}
The solution is
\begin{eqnarray}
f^1&=&f^2=0,\nonumber\\
a&\rightarrow&\infty.
\end{eqnarray}
This solution gives $V_{{\rm eff}}=0$ and {\it N}=2 supersymmetry seems to be
unbroken. However, such a solution is ruled out, since
$a>\Lambda$ is not allowed by the quantum deformation of the moduli space.
Therefore we conclude that there is no stable vacuum in the LEEA of SQED
with FI term under the assumption of no confinement\footnote{Without
FI term there is a stable vacuum, of course. We can define a theory on 
a point in the moduli space except for a=0 and $a>\Lambda$.
For small $a$, the LEEA reduces to the one obtained in the perturbation theory.}
.

\begin{figure}
 \psfig{file=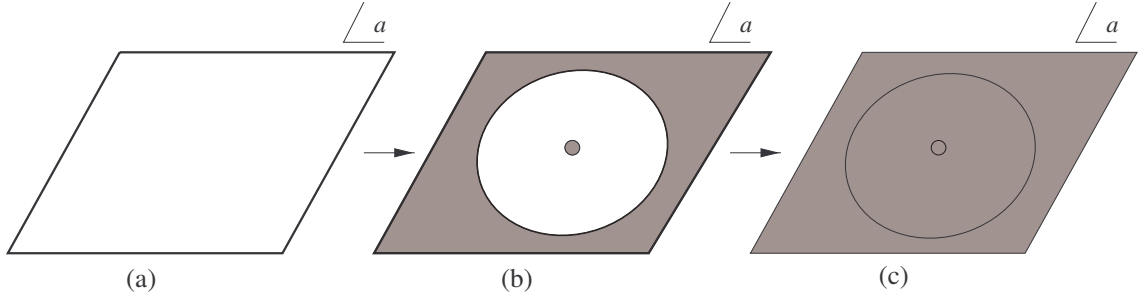,width=16.5cm}
\caption{(a) Classical modulispace: No constraint for $a$. (b) Quantum
 moduli space: Shaded region ($a=0,a>\Lambda$) denotes the excluded
 region. (c) Quantum moduli with
 FI term: No allowed region. For clearly visualizing the situation, we
 describe these pictures as if $a$ takes a complex value, though its
 phase is unphysical.}
\label{clas}
\end{figure}

Here, we summarize how the moduli space has been deformed.
In the classical theory the moduli space is
parameterized by $a$, and any value of $a$ is possible
(fig.\ref{clas}(a)). By the quantum effect the region $a>\Lambda$ and
a point $a=0$ are forbidden (fig.\ref{clas}(b)). By including FI term
remaining moduli space is lifted up and slopes down to $a=0$ axis, and
no stable vacuum exists (fig.\ref{clas}(c)).

We interpret this results as follows. Recall that
we assume that the confinement does not occur at low energies. 
Thus, the result no stable vacuum in the LEEA of SQED with FI term, suggests that the
confinement may occur at low energies. If we assume confinement at low
energies, we may be able to remove
the singularity at $a=0$ and may obtain a stable vacuum.

\begin{figure}
 \psfig{file=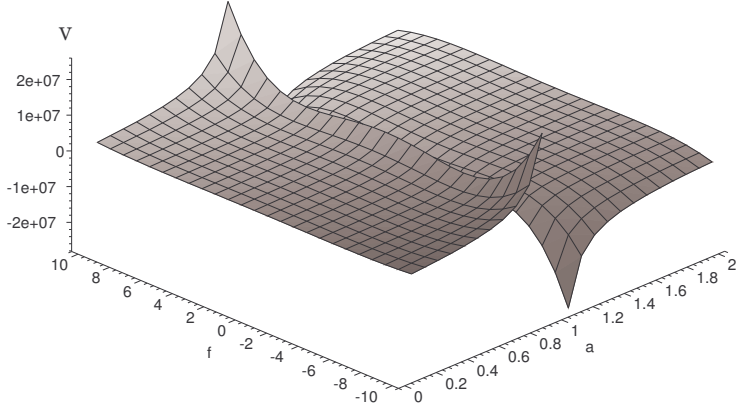,width=10cm}
\caption{The shape of the potential in the condition of Re$f^1={\rm Re}f^2={\rm Im} f^1={\rm Im} f^2=f$, $\Lambda^\prime=1$
 and $\xi^{(a)}=100$. }
\label{rittai}
\end{figure}

\begin{figure}
 \psfig{file=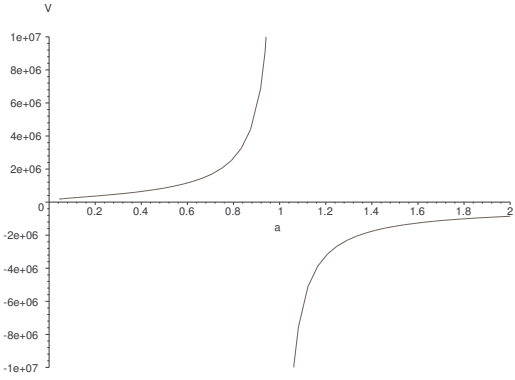,width=7cm}
\caption{The slice of the potential along $f=0$ axis.}
\label{danmen}
\end{figure}

The shape of the scalar potential is given in fig.\ref{rittai}. The
vacuum energy incorrectly takes negative value for
$a>\Lambda^\prime$. For 
$a<\Lambda^\prime$, the potential slopes down to $a=0$ axis where the
theory can not be
defined. The slice of the potential along
$f=0$ axis is shown in fig.\ref{danmen}. We have almost the same form
as in fig.\ref{danmen} for any slice along $f\neq 0$.

\begin{figure}
 \psfig{file=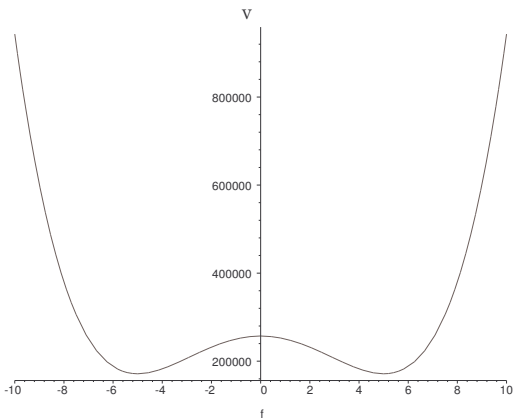,width=7cm}
\caption{The slice of the potential along $a=0.1$ with $\xi^{(a)}=100$.}
\label{danmen2}
\end{figure}

The structure along the axis of the constant $a$ is a little
complicated. We can understand it by referring the masses of two fields
$f^i$. They are obtained as
\begin{eqnarray}
m^2_{f_1,f_2}=(\sqrt{2}a)^2\pm\displaystyle\frac{2\pi^2}{\ln\frac{\Lambda}{a}}\left\{\sum_{a=1}^3(\xi^{(a)})^2\right\}^{\frac{1}{2}}.
\end{eqnarray}
Note that one of the squared masses can become negative for small value of 
$a$ satisfying condition 
\begin{eqnarray}
\displaystyle 2\pi^2\left\{\sum_{a=1}^3(\xi^{(a)})^2\right\}^{\frac{1}{2}}>(\sqrt{2}a)^2 \ln\frac{\Lambda^\prime}{a}.\label{eq:cond}
\end{eqnarray}
Fig.\ref{danmen2} shows the typical shape of the slice along $a\neq 0$ 
for small $a$.

\section{Conclusion}
To obtain some exact results of QED, we constructed the LEEA of SQED
with single massless matter including FI term. We assumed that the
confinement does not occur at low energies and the LEEA is described by
elementary fields. We found that the harmonic
superspace formalism is
very useful for applying symmetry and holomorphy in the construction. We
reproduced the LEEA of the chiral
part which is coincide with the result given by Seiberg and Witten. We
constructed the LEEA 
of the analytic part including FI term. This part was the tree-level
exact. We wrote down
the scalar potential of the LEEA and analyzed it.
We found that there is no stable vacuum, and could not define
the theory. We interpret this result as an evidence of the confinement
at low energies in non-supersymmetric QED.
If we assume there is confinement at low
energies, we may get rid of the singularity at $a=0$ and obtain a
stable vacuum. 

\appendix
\section{}
\noindent
Metric and anti-symmetric tensors:
\begin{eqnarray}
& g_{\mu\nu}=\mbox{\rm diag}(1,-1,-1,-1), &\\
& \epsilon_{\alpha\beta}= i\sigma_2=\left(
	\begin{array}{cc}
            0 & -1 \\
            1 & 0  
	\end{array}
        \right) ,\hspace{3mm} 
  \epsilon^{\alpha\beta}=-i\sigma_2=\left(
	\begin{array}{cc}
	    0 & 1 \\
	   -1 & 0 
	\end{array}
	\right), & \\
& \epsilon^{0123}=1,\hspace{3mm}\epsilon_{0123}=-1, &\\
& \psi^\alpha=\epsilon^{\alpha\beta}\psi_\beta,\hspace{3mm}\psi_\alpha=\epsilon_{\alpha\beta}\psi^\beta. &\\
\end{eqnarray}
Pauli matrices:
\begin{eqnarray}
& \sigma^0=\left(
	\begin{array}{cc}
		-1 &  0 \\
		0  & -1
	\end{array}
	\right) ,\hspace{3mm}
  \sigma^1=\left(
	\begin{array}{cc}
		0 & 1\\
		1 & 0
	\end{array}
	\right) ,\hspace{3mm}
  \sigma^2=\left(
	\begin{array}{cc}
		0 & -i\\
		i & 0
	\end{array}
	\right) ,\hspace{3mm}
  \sigma^3=\left(
	\begin{array}{cc}
		1 & 0 \\
		0 & -1 
	\end{array}
	\right). &
\end{eqnarray}
Supersymmetry algebra in the massless case:
\begin{eqnarray}
& \{Q_{i\alpha},\bar{Q}^j_{\dot{\alpha}}\}=-2i\delta_i^j\sigma^\mu_{\alpha\dot{\alpha}} P_\mu,\hspace{5mm}P_\mu=i\partial_\mu, &\\
& \{Q_{i\alpha},Q_{j\beta}\}=0, & \\
&\{\bar{Q}^i_{\dot{\alpha}},\bar{Q}_{\dot{\beta}}^j\}=0. &
\end{eqnarray}
Covariant derivatives in CB:
\begin{eqnarray}
&D_\alpha^i=\displaystyle\frac{\partial}{\partial\theta_i^\alpha}+i\sigma^\mu_{\alpha\dot{\alpha}}\bar{\theta}^{i\dot{\alpha}}\partial_\mu,&\\
&\bar{D}_{i\dot{\alpha}}=-\displaystyle\frac{\partial}{\partial{\bar{\theta}}^{i\dot{\alpha}}}-i\theta^\alpha_i\sigma^\mu_{\alpha\dot{\alpha}}\partial_\mu, &\\
&D^{++}=\displaystyle u^{+i}\frac{\partial}{\partial u^{-i}},&\\
&D^{--}=\displaystyle u^{-i}\frac{\partial}{\partial u^{+i}}.&
\end{eqnarray}
Covariant derivatives in AB:
\begin{eqnarray}
&\displaystyle D^+_{\alpha}=\frac{\partial}{\partial \theta^{-\alpha}},\hspace{3mm}\bar{D}^+_{\dot{\alpha}}=\frac{\partial}{\partial \bar{\theta}^{-\dot{\alpha}}},&\\
&\displaystyle D^-_{\alpha}=-\frac{\partial}{\partial \theta^{+\alpha}}+2i\sigma_{\alpha\dot{\alpha}}^\mu\bar{\theta}^{-\dot{\alpha}}\partial_\mu,&\\
&\displaystyle \bar{D}^-_{\dot{\alpha}}=-\frac{\partial}{\partial \bar{\theta}^{+\dot{\alpha}}}-2i\theta^{-\alpha}\sigma_{\alpha\dot{\alpha}}^\mu\partial_\mu,&\\
&D^{++}=\displaystyle u^{+i}\frac{\partial}{\partial u^{-i}}-2i\theta^+\sigma^\mu\bar{\theta}^+\frac{\partial}{\partial x^\mu_A}+\theta^{+\alpha}\frac{\partial}{\partial \theta^{-\alpha}}+\bar{\theta}^+_{\dot{\alpha}}\frac{\partial}{\partial \bar{\theta}^{-}_{\dot{\alpha}}},\label{eq:cov}&\\
&D^{--}=\displaystyle u^{-i}\frac{\partial}{\partial u^{+i}}-2i\theta^-\sigma^\mu\bar{\theta}^-\frac{\partial}{\partial x^\mu_A}+\theta^{-\alpha}\frac{\partial}{\partial \theta^{+\alpha}}+\bar{\theta}^{-}_{\dot{\alpha}}\frac{\partial}{\partial \bar{\theta}^{-}_{\dot{\alpha}}}.&
\end{eqnarray}
Some useful algebras:
\begin{eqnarray}
&\{D_\alpha^+,D_\beta^-\}=0,
\hspace{2mm}\{D^+_{\alpha},\bar{D}_{\dot{\beta}}^- \}=-2i\sigma^\mu_{\alpha\dot{\beta}}\partial_\mu,& \\
&\{\bar{D}^+_{\dot{\alpha}},D_\beta^-\}=2i\sigma^\mu_{\beta\dot{\alpha}}\partial_\mu,\hspace{2mm}
\{ \bar{D}_{\dot{\alpha}}^+,\bar{D}_{\dot{\beta}}^- \}=0,&\\
&[D_\alpha^+,D^{--}]=-D_\alpha^-,\hspace{2mm}[\bar{D}_{\dot{\alpha}}^+,D^{--}]=-\bar{D}_{\dot{\alpha}}^-,& \\
&[D^-_{\alpha},D^{++}]=-D^+_\alpha,\hspace{2mm}[\bar{D}^-_{\dot{\alpha}},D^{++}]=-\bar{D}^+_{\dot{\alpha}},&\\
&\left[D^{++},D^{--}\right]=D^0=\displaystyle u^{+i}\frac{\partial}{\partial u^{+i}}-u^{-i}\frac{\partial}{\partial u^{-i}}.&
\end{eqnarray}

\newpage
\begin{table}
\begin{center}
\begin{tabular}{cccc}
          & U(1$)_{{\rm g}}$ & U(1) & U(1$)_{{\rm R}}$ \\\hline
$\phi^+$  &  1               &  1   & 0        \\
$V^{++}$  &  $-$               &  2   & 0        \\
$W$       &  0               &  0   & 2        \\
$\Lambda$ &  0               &  0   & 2       
\end{tabular}
\caption{The transformations laws in the fundamental theory. U(1$)_{{\rm g}}$ denotes gauge group and U(1)
 denotes the projected charge of the global symmetry SU(2$)_{{\rm R}}$. The scale of dynamics $\Lambda$ is
 a parameter of the theory. We assign  U(1$)_{{\rm R}}$ charge 2 to it,
 by which the theory has non-anomalous U(1$)_{{\rm R}}$ symmetry.}
\label{sym1}
\end{center}
\end{table}

\begin{table}
\begin{center}
\begin{tabular}{cccc}
           & U(1$)_{\rm g}$ & U(1) & U(1$)_{\rm R}$ \\\hline
 $\phi^+$  &        1       &   1  &        0      \\
 $V^{++}$  &       $-$      &   2  &        0      \\
 $W$       &        0       &   0  &        2      \\
 $\Lambda$ &        0       &   0  &        2      \\      
 $\xi^{++}$&       0       &   2  &        0
\end{tabular}
\caption{The transformation laws in the fundamental theory with FI term.}
\label{sym2}
\end{center}
\end{table}

\end{document}